\documentclass[oneside]{article}

\usepackage[utf8]{inputenc}
\usepackage[a4paper, margin=4cm]{geometry}
\usepackage{parskip}
\usepackage{comment}
\usepackage{amsmath} 
\usepackage{amssymb}
\usepackage{enumerate}
\usepackage[english]{babel}
\usepackage{amsmath}
\usepackage{wasysym}
\usepackage{listings}
\usepackage{url}
\usepackage{mathtools}
\usepackage{amsthm}
\usepackage{cleveref}
\usepackage{multirow}
\usepackage{subfig,graphicx}
\usepackage[toc,page,title]{appendix}
\PassOptionsToPackage{table}{xcolor}
\usepackage{colortbl}
\usepackage{booktabs}
\usepackage[table]{xcolor}
\usepackage{array}
\usepackage{caption}
\usepackage[export]{adjustbox}
\usepackage[authoryear]{natbib}
\usepackage{float}
\floatstyle{plaintop}
\restylefloat{table}

\numberwithin{equation}{section}

\begin{document}

\title{Vegetation Patterning Can Both Impede and Trigger Critical Transitions from Savanna to Grassland
\author{
Jelle van der Voort\thanks{Mathematical Institute, Leiden University, 2300 RA, Leiden, Netherlands} 
\and 
Mara Baudena\thanks{National Research Council, Institute of Atmospheric Sciences and Climate (CNR-ISAC), 10133, Torino, Italy; National Biodiversity Future Center (NBFC), 90133, Palermo, Italy} 
\and 
Ehud Meron\thanks{The Swiss Institute for Dryland Environmental and Energy Research, BIDR, Ben-Gurion University of the Negev, Midreshet Ben-Gurion 8499000, Israel; Physics Department, Ben-Gurion University of the Negev, Beer-Sheva 8410501, Israel} 
\and 
Max Rietkerk\thanks{Copernicus Institute of Sustainable Development, Section Environmental Sciences, Utrecht University, Utrecht, the Netherlands} 
\and 
Arjen Doelman$^*$
}
}

\maketitle

\textbf{Abstract}

Tree-grass coexistence is a defining feature of savanna ecosystems, which play an important role in supporting biodiversity and human populations worldwide. While recent advances have clarified many of the underlying processes, how these mechanisms interact to shape ecosystem dynamics under environmental stress is not yet understood. Here, we present and analyze a minimalistic spatially extended model of tree-grass dynamics in dry savannas. We incorporate tree facilitation of grasses through shading and grass competing with trees for water, both varying with tree life stage. Our model shows that these mechanisms lead to grass-tree coexistence and bistability between savanna and grassland states. Moreover, the model predicts vegetation patterns consisting of trees and grasses, particularly under harsh environmental conditions, which can persist in situations where a non-spatial version of the model predicts ecosystem collapse from savanna to grassland instead (a phenomenon called ``Turing-evades-tipping''). Additionally, we identify a novel ``Turing-triggers-tipping'' mechanism, where unstable pattern formation drives tipping events that are overlooked when spatial dynamics are not included. These transient patterns act as early warning signals for ecosystem transitions, offering a critical window for intervention. Further theoretical and empirical research is needed to determine when spatial patterns prevent tipping or drive collapse.

\vspace*{\fill}

Author for correspondence: Jelle van der Voort\\
Email: \textit{j.van.der.voort@math.leidenuniv.nl} 

\newpage

\section{Introduction}

The coexistence of trees and C4 grasses defines savannas, which support a significant proportion of the global biodiversity and human populations \citep{Sankaran2005,Parr2014}. The long-standing question of how trees and grasses persist together, despite competing for the same limited resources like water, is essential for assessing the existence, stability and resilience to environmental change of savanna ecosystems. A recent review \citep{Holdo2023} clarifies the current understanding of the coexistence problem, distinguishing between mesic savannas (mean annual precipitation $> 650$ mm) and dry savannas ($< 650$ mm) \citep{Sankaran2005, DOnofrio2018, Holdo2018}. In mesic savannas, tree-grass balance is maintained primarily by disturbances such as herbivory and fire, while in dry savannas, water stress plays a dominant role. Despite these advances, recent models overlook key dynamics specific to dry savannas, particularly the nature of tree-grass interactions including their spatially extended structure.  

Empirical studies of dry savannas consistently show that grasses exert significant competitive pressure on trees \citep{Riginos2009, February2013}, especially hindering establishment, while trees seedlings have minimal competitive impact on grasses \citep{February2013, Campbell2017,Holdo2023}. Despite this, trees persist due to their deeper rooting and flexible water-use strategies, enabling coexistence with shallow-rooted grasses that dominate topsoil water acquisition \citep{Nippert2007, Kulmatiski2013, Holdo2023}. This supports the foundational hypothesis of niche partitioning along a soil depth axis \citep{Walter1976}. Despite their resilience, trees in dry savannas cannot develop sufficient leaf area to form closed canopies due to limited rainfall, preventing them from effectively shading out grasses \citep{Sankaran2005, Lehmann2011, DOnofrio2018, Holdo2023}. This limitation is captured by the ``Sankaran curve", which represents the existence of an upper boundary to tree cover in low-precipitation regions. Under such water-limited conditions, tree shading can actually alleviate moisture stress for grasses, facilitating their growth \citep{Scholes1997, Holdo2023}. For instance, in Kruger National Park, trees increased grass biomass beneath their canopies under low precipitation rates \citep{Moustakas2013}. This aligns with the stress-gradient hypothesis, which suggests that facilitation becomes more important as environmental stress increases \citep{Callaway1997,Dohn2013, Hernandez2022}.

These local dynamics create an asymmetrical interaction where grasses strongly compete with trees for water in the upper soil layers, particularly hindering tree establishment \citep{February2013,Campbell2017,Holdo2023}. However, as trees mature and develop deeper roots, they begin to facilitate grass growth by providing shade. Despite clear experimental evidence supporting both facilitation and competition across tree life stages, current savanna models lack accurate representations that capture both these dynamics and spatial aspects, such as seed dispersal \citep{Sankaran2004, Riginos2009, Holdo2023}.

\begin{figure}[t!]
    \centering
    \begin{minipage}[b]{0.5\textwidth}
        \centering
        \includegraphics[width=\textwidth]{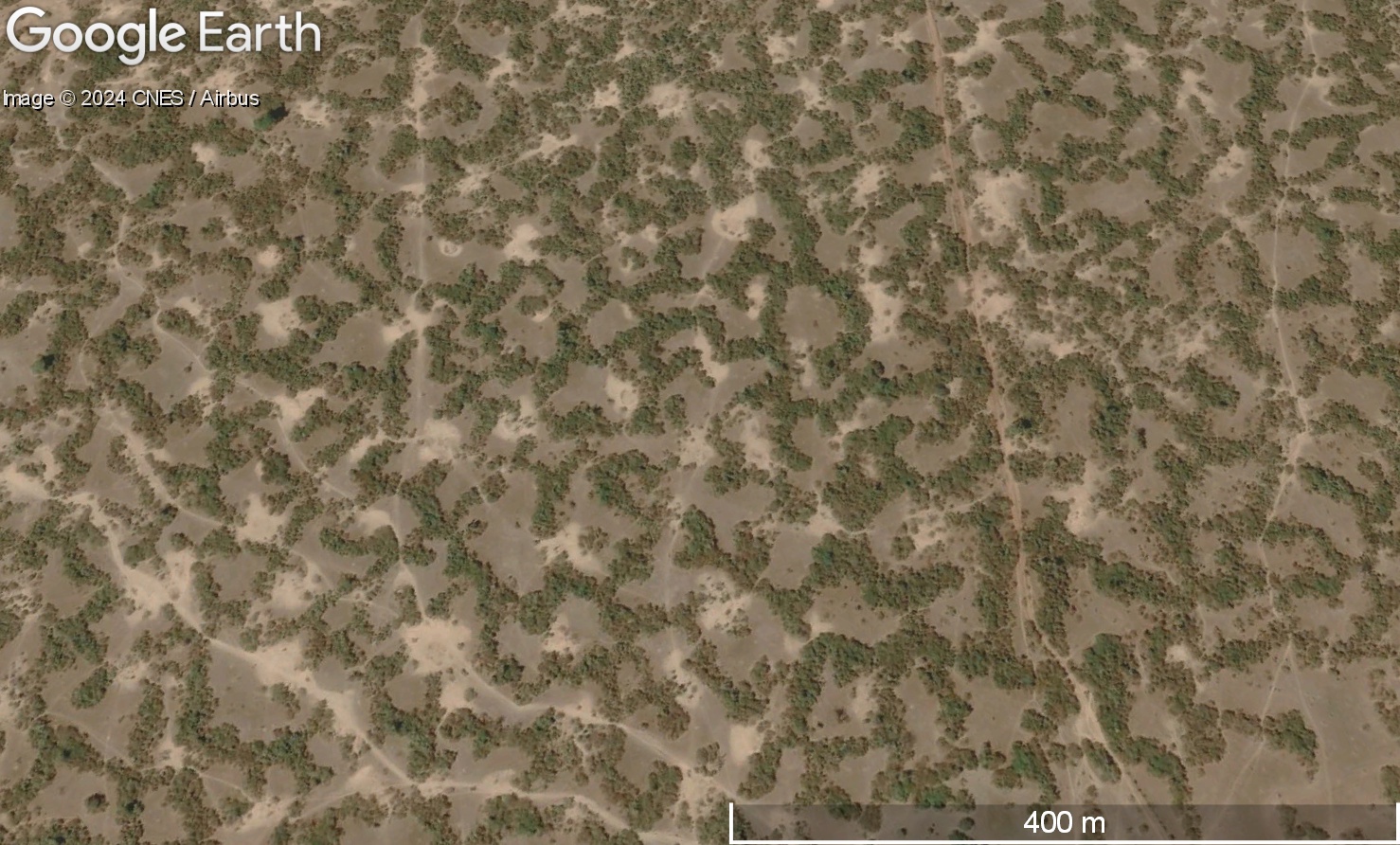}
    \end{minipage}%
    \hspace{0pt}%
    \begin{minipage}[b]{0.5\textwidth}
        \centering
        \includegraphics[width=\textwidth]{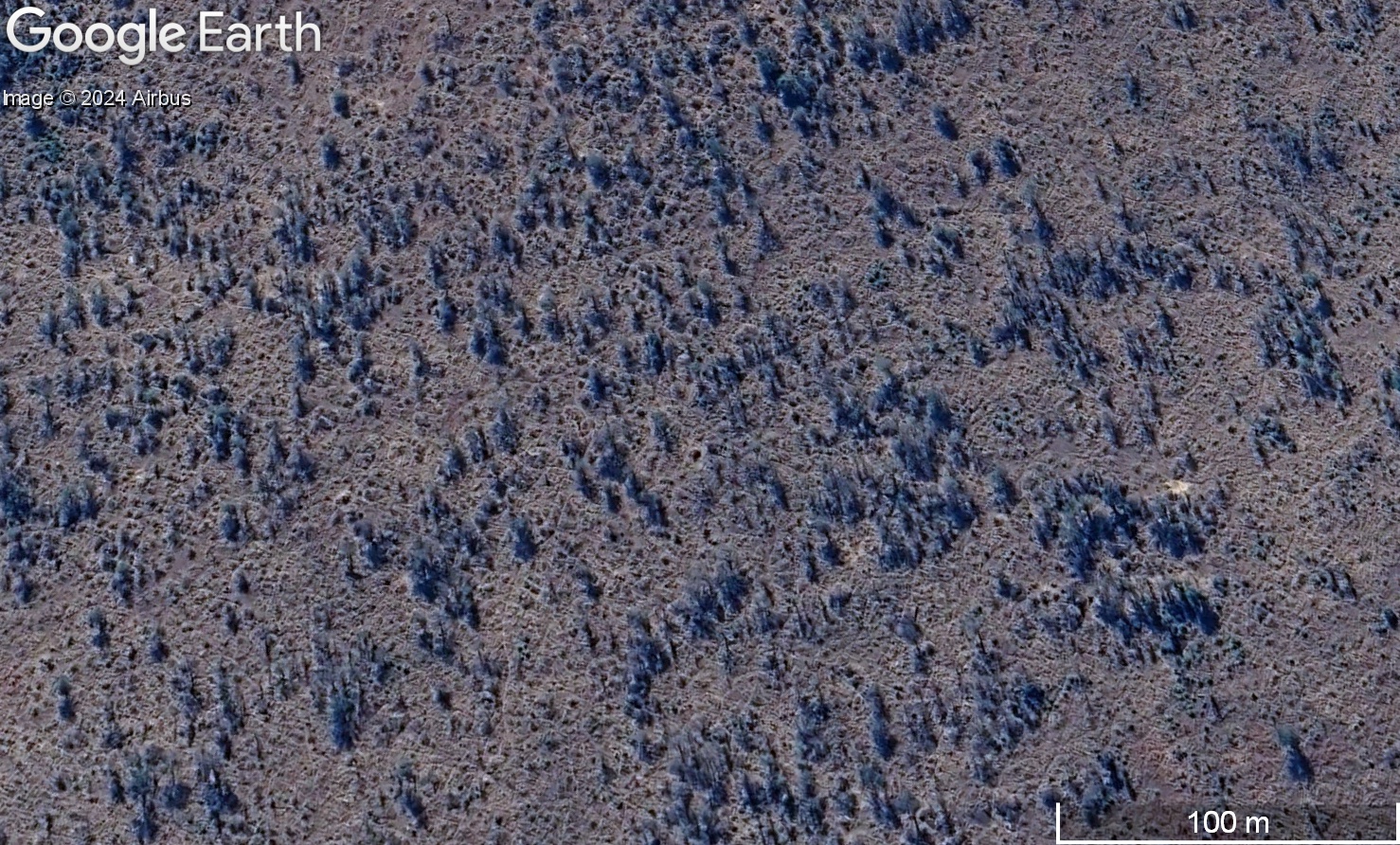}
    \end{minipage}
    \caption{Aerial photographs of vegetation self-organization in dry savannas. (left) Labyrinth patterns, Mali (13°03'15"N 6°40'51"W), via \citep{Groen2007}, (right) clustering of trees, South Africa (25°02'03"S 31°37'18"E). Pictures via Google Earth \citep{GoogleEarth}.}
    \label{fig:aerial_grid}
\end{figure}

Including spatial dynamics is crucial, as these interactions inherently influence the stability and resilience of savanna ecosystems under environmental change \citep{Rietkerk2021}. Without considering spatial effects, we overlook the widespread phenomenon of vegetation self-organization \citep{Staver2018} (Figure \ref{fig:aerial_grid}). Understanding these spatial patterns, a particular form of tree-grass coexistence, is essential for assessing ecosystem responses to environmental change, especially near tipping points \citep{Banerjee2024}. In this context, tipping refers to a sudden transition to an alternative state characterized by different ecosystem services. For instance, while grasslands provide important services like grazing, a shift from savanna to grassland results in the loss of tree-related services like shading and wildlife habitat.

Vegetation patterns are often viewed as early warning signals of ecological tipping points that may lead to ecosystem collapse \citep{Rietkerk2004}. Simultaneously, pattern formation may also enhance ecosystem resilience by delaying the onset of such tipping points \citep{Rietkerk2021}. For instance, spatially explicit models reveal that patterns can remain stable even under parameter changes that would induce collapse in corresponding non-spatial models \citep{Lefever1997,vonHardenberg2001,Rietkerk2002}. This phenomenon, which we refer to as ``Turing-evades-tipping” \citep{Rietkerk2021}, emphasizes the the significance of spatial interactions in enhancing ecosystem resilience and underscores the limitations of non-spatial approaches \citep{Banerjee2024}. Various spatial models have successfully captured vegetation patterning and pattern dynamics across diverse ecosystems \citep{Klausmeier1999,Hille2001,Rietkerk2002,Gilad2007,Zelnik2015,Getzin2016,Bastiaansen2018,Bennett2023II,Bennett2023}, including regular tree-grass patterning in a savanna environment \citep{Tega2022} and shifts from competition to facilitation consistent with the stress-gradient hypothesis \citep{Gilad2007II}. However, these models generally lack life-stage specific dynamics, a critical component in savanna ecosystems.

This study presents a conceptual spatially extended modelling framework to investigate tree-grass dynamics in dry savannas, drawing inspiration from the extensive review by \citep{Holdo2023}. We incorporate key recommendations from their work, including asymmetry in tree-grass interactions across different tree life stages: grasses inhibit the establishment of early-life stage trees, while mature trees facilitate grasses through shading. To simplify the mathematical analysis, we adopt a two-variable approach (trees and grasses), that captures the effects of resource limitation (soil moisture) implicitly, rather than explicitly modelling the resource itself. Additionally, instead of modelling trees as individuals, we represent them as biomass densities at the community level, allowing us to capture tree dynamics across life stages within a single equation while maintaining a manageable dynamical systems approach.

We analyze both the non-spatial version of the model in terms of ordinary differential equations and its spatially extended counterpart in terms of partial differential equations to investigate tipping phenomena. This approach allows us to study the existence and stability of tree-grass coexistence states, and the resilience to environmental change of dry savannas including the asymmetry of tree-grass interactions at different tree life stages. Specifically, we aim to answer the following questions: Can the interplay of facilitation and competition, varying with tree life stage, support stable tree-grass coexistence? How do spatial interactions modify the dynamics predicted by the non-spatial model? And what role does tree-grass patterning play in shaping ecosystem resilience under environmental change?

\section{Methods}

\subsection*{Modelling framework}

\subsubsection*{The model}
We propose a reaction-diffusion model given by the following equations,
\begin{align}
\begin{split}
    \frac{\partial g}{\partial t} &= a_g g \left(1-\frac{g}{K_g}\right) - m_g g + p_f g\alpha(s) + d_g \Delta g,\\
    \frac{\partial s}{\partial t} &= a_s s\left(\frac{s}{C_s}-1\right)\left(1-\frac{s}{K_s}\right) - m_s s - p_c gs\beta(s) + d_s \Delta s.
\end{split}
\label{eqn: the model}
\end{align}

Here, $g = g(\mathbf{x},t)$ and $s= s(\mathbf{x},t)$ represent the grass and savanna tree biomass densities, respectively, in kg m\textsuperscript{-2} at location $\mathbf{x}$ and time $t$. Our model assumes logistic growth for grass, an Allee-effect growth function for trees, facilitation of grasses by adult trees through shading (which reduces evaporation), competition for soil moisture between grasses and early-life-stage trees, and spatial effects introduced through seed dispersal modeled as diffusion. The model parameters and their typical ranges are presented in Table \ref{tab: GT model}. Parameter values are based on empirical field data where available and calibrated otherwise (see Appendix \ref{parametrization and calibration} for details).

\begin{table}[h!]
\centering
\arrayrulecolor{black}
\begin{tabular}{|>{\centering\arraybackslash}m{2cm}|>{\centering\arraybackslash}m{5.6cm}|>{\centering\arraybackslash}m{2cm}|>{\centering\arraybackslash}m{1.5cm}|}
\hline
\rowcolor[HTML]{EFEFEF} 
\textbf{Parameter} & \textbf{Description} & \textbf{Range} & \textbf{Units} \\ \hline
$a_g$  & Growth rate of grass & $1.5-5.1$ & yr\textsuperscript{-1} \\ \hline
$K_g$  & Theoretical grass biomass cap & $0.2-2.0$ & kg m\textsuperscript{-2} \\ \hline
$m_g$  & Mortality rate of grass & $1.5-4.6$ & yr\textsuperscript{-1} \\ \hline
$p_f$  & Rate of facilitation & $0-2$ & yr\textsuperscript{-1} \\ \hline
$d_g$  & Diffusion rate of grass  & $50-100$ & m\textsuperscript{2} yr\textsuperscript{-1} \\ \hline
$a_s$  & Growth rate of trees & $0.001-0.1$ & yr\textsuperscript{-1} \\ \hline
$C_s$  & Threshold tree density  & $0 - K_s/100$ & kg m\textsuperscript{-2}   \\ \hline
$K_s$  & Theoretical tree biomass cap & $3-10$ & kg m\textsuperscript{-2}  \\ \hline
$m_s$  & Mortality rate of trees & $0.046 - 0.46$ & yr\textsuperscript{-1} \\ \hline
$p_c$  & Rate of competition & $5.99-34.5$ & yr\textsuperscript{-1} \\ \hline
$d_s$ & Diffusion rate of trees & $5-10$ & m\textsuperscript{2} yr\textsuperscript{-1} \\ \hline
\end{tabular}
\caption{Description of the model parameters together with realistic ranges, based on parametrization from literature and model calibration (see Appendix \ref{parametrization and calibration}).}
\label{tab: GT model}
\end{table}

\subsubsection*{Vegetation growth}

Grass growth is modelled using a logistic function, reflecting how grasses expand toward a maximum standing biomass, typical of the continuous grass canopy in dry savannas despite resource constraints like water \citep{Holdo2023}. Assuming $a_g>m_g$ excludes bare soil stability, aligning with a focus on savanna ecosystems where grasses are relatively abundant. The model can be adapted in a straightforward manner to include bare soil dynamics by replacing the logistic growth function with an Allee-effect growth function, mimicking vegetation-water feedbacks observed in existing dryland models \citep{Klausmeier1999,Rietkerk2002,Bennett2023II}.

In contrast, savanna trees are more sensitive to water stress, particularly during early stages of development where they struggle due to their shallow root systems, competing with grasses for water in the topsoil layer. Once matured, trees become more resilient to drought, accessing deeper soil moisture beyond the reach of grasses and exhibiting accelerated growth, even under harsh conditions \citep{Nippert2007,Kulmatiski2013,Holdo2023}. To capture these essential aspects related to water limitation, we apply an Allee-effect growth function. This function reflects both the rooting advantages of mature trees (i.e., large biomass density values), and the shading benefits they provide, which reduce evaporation and enhance tree survival \citep{Ma2023}. In contrast, early-life stage trees (i.e., small biomass density values), limited to shallower roots and lacking sufficient leaf area for shading, are more vulnerable to water stress. Natural mortality is modelled as linear decay term for both trees and grasses.

\subsubsection*{Interaction terms}

\begin{figure}[b!]
    \centering
    \begin{adjustbox}{center, margin=0cm}
  \includegraphics[width=0.6\textwidth]{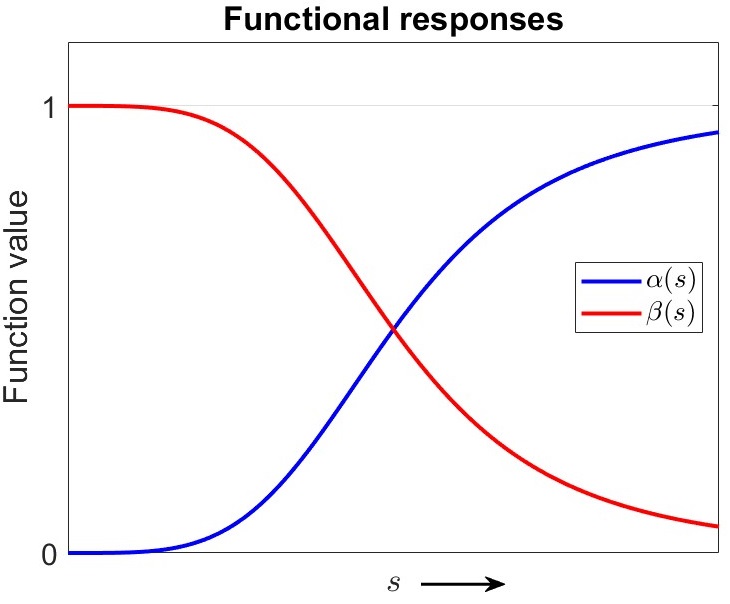}
\end{adjustbox}
\caption{The facilitative effect of trees on grasses, $\alpha(s)$, and the diminishing competitive pressure of grasses on trees, $\beta(s)$, both shown as functions of increasing tree density.}
\label{fig: Tree-Grass Interaction Dynamics}
\end{figure}

Mature savanna trees alleviate moisture stress for grasses through shading, creating a facilitative effect \citep{Scholes1997,Moustakas2013}. The magnitude of this benefit increases with tree size, with smaller or younger trees providing minimal shade. Therefore, at low tree biomass densities, the shading effect is negligible. We model this facilitation using the function $\alpha(s)$, representing the positive impact of tree shading on grasses as tree density increases (Figure \ref{fig: Tree-Grass Interaction Dynamics}). At even larger tree biomass densities, facilitation may diminish and transition into competition for light and other resources \citep{Holdo2023}. However, water scarcity in dry savannas typically prevents the formation of closed canopies \citep{Sankaran2005,Lehmann2011,DOnofrio2018}, limiting the extent of this competitive effect. Thus, we focus on conditions where facilitation remains dominant.

In their early-life stages, savanna trees establish roots in the upper soil layers where grasses limit water access for juvenile trees, inhibiting their growth and establishment \citep{Holdo2023}. This competitive inhibition is modelled by the term $-p_cgs\beta(s)$, which represents an additional mortality factor at low tree biomass densities, reflecting the intense competition young trees face from grasses. As trees mature and develop deeper roots, they access water unavailable to grasses, reducing this competitive pressure. The function $\beta(s)$ captures this shift, inducing that grass competition remains strong at low tree biomass but decreases as tree biomass increases (Figure \ref{fig: Tree-Grass Interaction Dynamics}).

\subsubsection*{Seed dispersal and biomass redistribution}
We model the spatial effects of seed dispersal using diffusion, a standard approach in spatially extended vegetation models \citep{Klausmeier1999, Hille2001, Gilad2007, Bennett2023}. In dry savannas, grasses propagate rapidly, forming a nearly continuous grass layer \citep{Holdo2023}. In contrast, savanna trees do not form closed canopies in these arid environments \citep{Sankaran2005}. Although tree seed dispersal rates may be comparable to those of grasses, tree seeds face greater challenges in germination, and newly emerged seedlings require favorable conditions to survive, resulting in a lower overall rate of biomass redistribution.  To capture these dynamics, we assign a higher diffusion coefficient to grasses ($d_g$) than to trees ($d_s$), as the diffusion coefficient reflects not only the rate of spatial seed spread but also the constraints on the redistribution of biomass imposed by germination and establishment. When spatial effects are excluded, i.e., when $d_g = d_s = 0$, the model simplifies to a non-spatial system. This non-spatial version is useful for studying the coexistence problem and provides a baseline for comparison. We refer to the spatially extended system as the ``spatial model" and the version without diffusion as the ``non-spatial model".

\subsubsection*{The role of water availability}
We account for water availability indirectly by modelling its effects through facilitation and competition between trees and grasses at different life stages. Tree shading reduces moisture stress on grasses, while grasses compete with trees for water in the upper soil layers. These interactions help us to explore whether variations in dispersal and establishment rates drive stable tree-grass coexistence and spatial patterns. To further examine the influence of water availability along rainfall gradients, which affects the relative dominance of grasses and trees \citep{Lehmann2011,DOnofrio2018}, we use the parameter $a_s$ (which governs the tree growth rate) as a proxy for water availability in the upcoming bifurcation analysis. While water availability likely affects multiple model parameters, including the grass growth rate $a_g$, we focus on $a_s$ because tree biomass accumulation is primarily limited by water stress in the lower range of the Sankaran curve \citep{Sankaran2005}. Moreover, varying a single parameter ensures a tractable bifurcation analysis, and we note that the bifurcation structure in our model remains qualitatively similar when multiple parameters are varied. This approach provides a valid approximation of the impact of water stress on vegetation patterns without explicitly modelling water as a dynamic variable.

\subsection*{Vegetation pattern formation theory}

To analyze vegetation pattern formation in dry savannas and its impact on ecosystem resilience, we will focus on detecting Turing bifurcations. These bifurcation drive the emergence of spatial patterns as environmental parameters cross a critical threshold \citep{Rietkerk2008,Meron2019}. Turing bifurcations can be classified as supercritical or subcritical, each with distinct ecological implications.

In a supercritical Turing bifurcation, small deviations from a uniform state result in the formation of stable, small-amplitude patterns. As the system moves away from the bifurcation point, pattern characteristics, such as amplitude and width, adapt continuously until the stability boundary (the edge of the so-called Busse balloon \citep{Bastiaansen2018,Rietkerk2021}) is crossed, where wave number adaptation occurs. This bifurcation is often considered non-catastrophic, inducing smooth transitions from uniform to patterned states and potentially enhancing ecosystem resilience \citep{Siteur2014}.

Conversely, a subcritical bifurcation leads to transient patterns with increasing amplitude that drive the system toward diverse ecological outcomes, including large-amplitude patterns with sharp boundaries, such as fronts and spikes \citep{Doelman2019,Tzuk2020}. These interacting localized structures emerge from scale separation, typically when species and resource spread rates differ significantly across the landscape. Multi-stability of such patterns, for instance homoclinic snaking \citep{Knobloch2008}, can also occur \citep{Zelnik2015,Meron2019}. Alternatively, transient patterning can guide the system towards a different uniform steady state, which may represent a form of ecosystem collapse if the steady states differ considerably. This transition is induced by bistability, often complicating recovery (hysteresis), making subcritical bifurcations potentially harmful.

Distinguishing between these bifurcation types is crucial for understanding how spatial patterns influence the stability and resilience of dry savannas under environmental stress.

\section{Results}

\subsection*{Analysis of the non-spatial model}

\begin{figure}[t!]
    \centering
    \begin{adjustbox}{center, margin=0cm}
  \includegraphics[width=\textwidth]{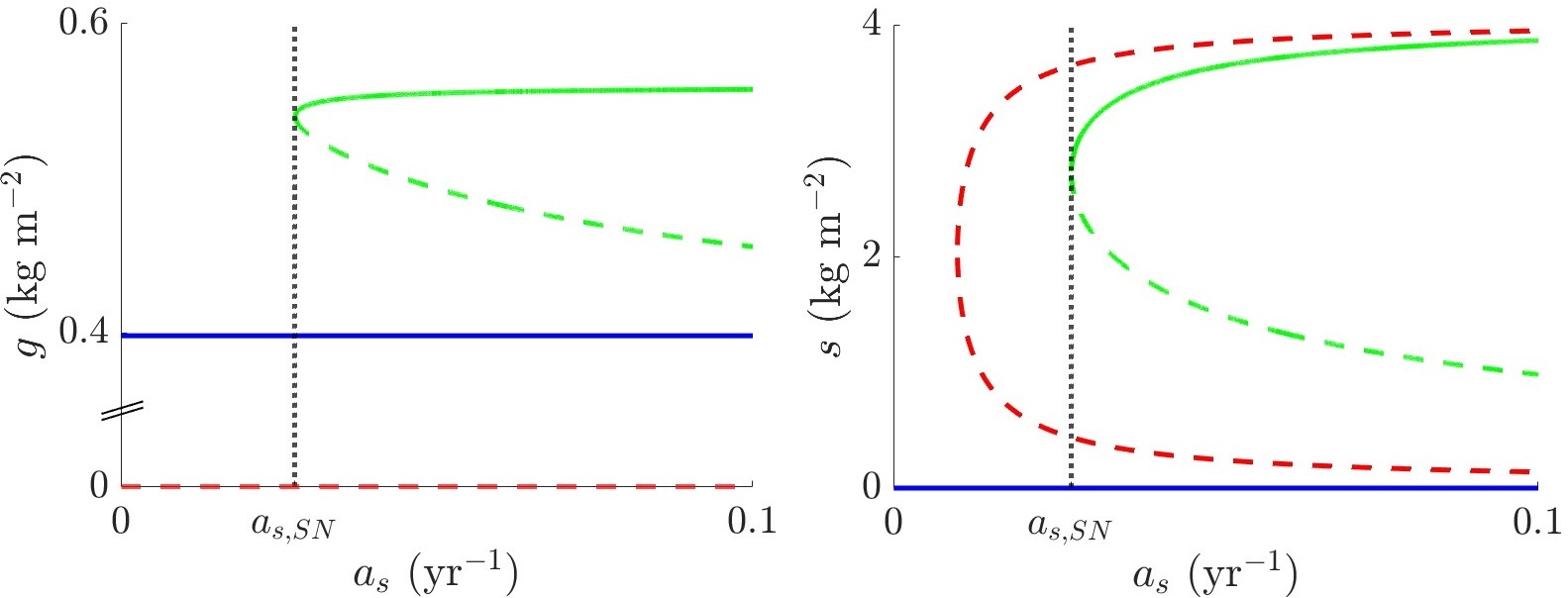}
\end{adjustbox}
\caption{Bifurcation diagrams for grass $g$ (left) and savanna trees $s$ (right) in the non-spatial model, plotted as a function of $a_s$, a proxy for water availability. The blue branch represent the grass-only state, the green branches the tree-grass coexistence states, and the red branches the tree-only states. Solid lines indicate stable equilibria and dashed lines indicate equilibria that are unstable. The diagrams highlight the bistability for $a_s > a_{s,SN}$, and the sudden transition from a savanna state to a grassland as the threshold $a_{s,SN}$ is crossed under increasing water stress.}
\label{Bifurcation diagram nonspatial}
\end{figure}

We start by analyzing the non-spatial model, focusing on stable tree-grass coexistence. The technical details of the linear stability analysis are provided in Appendix \ref{linear stability analysis}, while the key results are summarized below.

The non-spatial system admits multiple homogeneous equilibria, with the exact number depending on specific parameter setting. In all cases, both a bare soil equilibrium and a grass-only equilibrium are present. Given the parametrization, the bare soil state is unstable, while the grassland state is always stable, consistent with the assumption of logistic growth favoring grass persistence in dry savannas. Additionally, two unstable tree-only equilibria commonly arise from a saddle-node bifurcation, reflecting the inability of trees to close their canopy in dry savannas \citep{Sankaran2005}. Notably, the system allows for the existence of one stable equilibrium where both grass and trees coexist, representing a savanna state. 

For high values of $a_s$ (which governs the tree growth rate and acts as a proxy of water availability), two stable equilibria coexist: a grass-only state and a tree-grass coexistence state (Figure \ref{Bifurcation diagram nonspatial}). As $a_s$ decreases beyond the critical threshold $a_{s,SN}$, this bistability is lost in a saddle-node bifurcation, representing an ecological tipping point. This suggests that under slowly increasing water stress, ecosystems can abruptly shift from a savanna to a treeless grassland. Further bifurcation analysis indicates that parameters favoring trees when increased (i.e., $K_s$) enhance the range of tree-grass coexistence, while those that favor grasses (i.e., $a_g$, $K_g$, $p_f$) or negatively influence trees (i.e., $C_s$, $m_s$ and $p_c$) limit this range.

The existence of a stable tree-grass coexistence state in the non-spatial model underscores that the balance between facilitation and competition at different tree life stages can sustain tree-grass coexistence in dry savannas, even before accounting for spatial interactions. 

\subsection*{Vegetation patterning under environmental stress}

In analyzing the spatial model, we focus on parameter ranges where tree shading substantially alleviates moisture stress for grasses, a condition typical of dry savannas experiencing severe water limitation. This strong facilitative effect is represented by high values of $p_f$, within the ecologically relevant range provided in Table \ref{tab: GT model}.

\begin{figure}[t!]
    \centering
    \begin{adjustbox}{center, margin=0cm}
  \includegraphics[width=\textwidth]{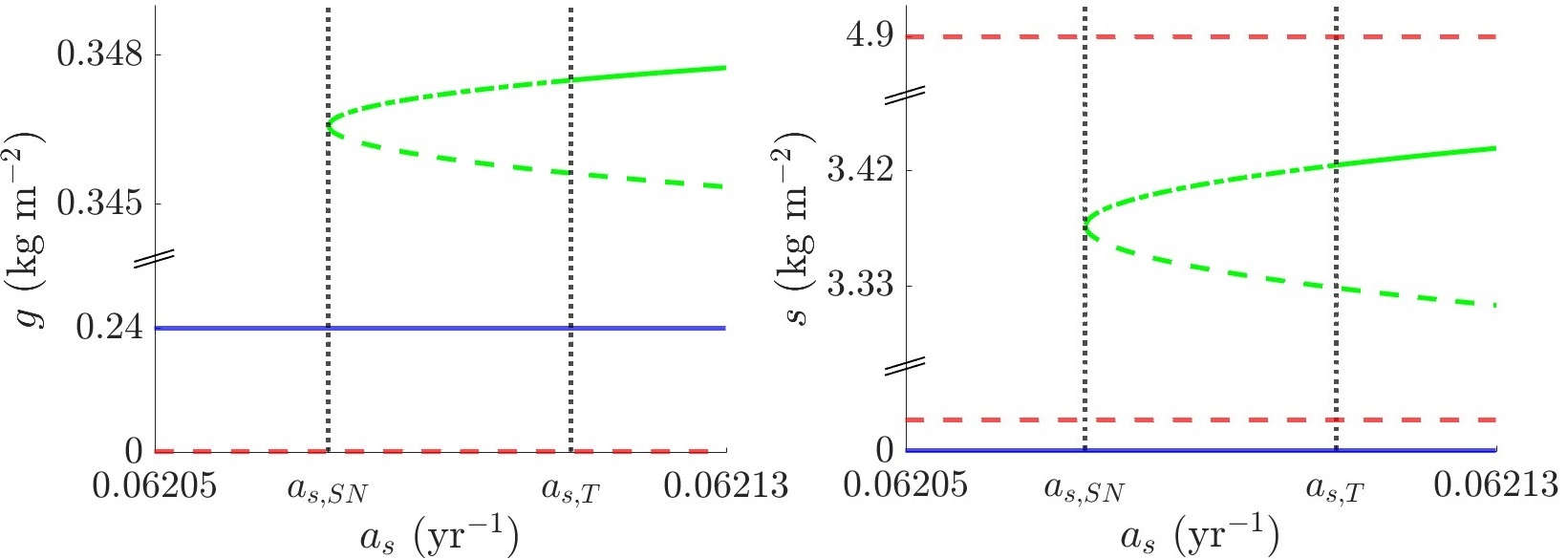}
\end{adjustbox}
\caption{Bifurcation diagrams for grass $g$ (left) and savanna trees $s$ (right) in the spatial model, plotted as a function of $a_s$, a proxy for water availability. The colors and the line styles have the same meaning as in Figure \ref{Bifurcation diagram nonspatial} with the addition of dash-dotted lines indicating equilibria stable to homogeneous perturbations but unstable to heterogeneous perturbations. The diagrams highlight the onset of spatial instability at $a_{s,T}$, which occurs at higher values than the saddle-node bifurcation at $a_{s,SN}$ that drives tipping in the non-spatial model.}
\label{Bifurcation diagram spatial}
\end{figure}

Within this parameter region, the uniform savanna state may undergo a Turing bifurcation, signaling spatial instability (see Appendix \ref{linear stability analysis} for the linear stability analysis and dispersion relation). The Turing bifurcation occurring at $a_{s,T}$ signals the onset of spatial patterning preceding the tipping point at $a_{s,SN}$ (Figure \ref{Bifurcation diagram spatial}). In our numerical experiments, this bifurcation is subcritical, generating patterns that grow in amplitude and act as transients until they approach another stable structure. 

\begin{figure}[b!]
    \centering
    \begin{adjustbox}{center, margin=0cm}
  \includegraphics[width=\textwidth]{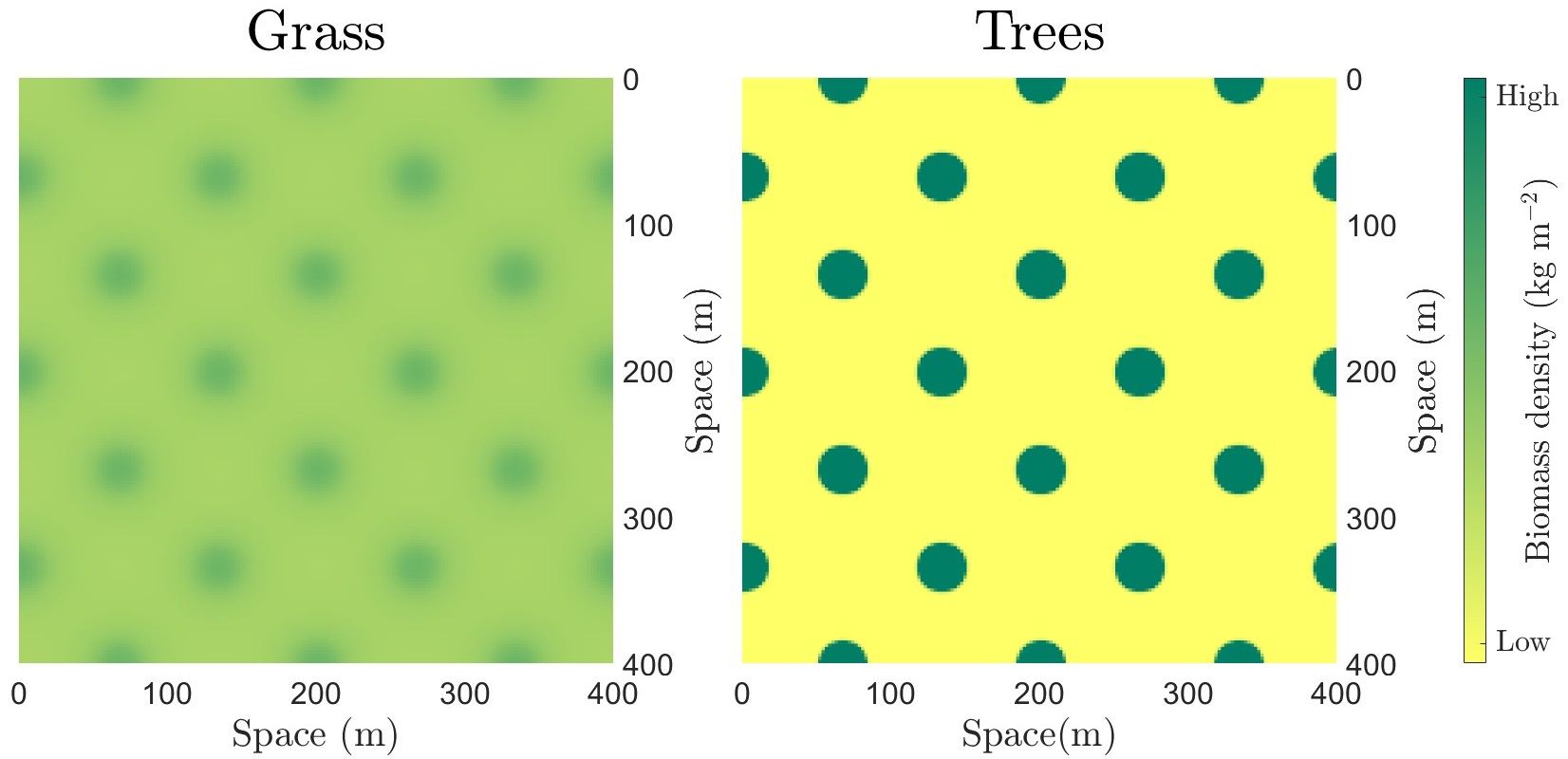}
\end{adjustbox}
\caption{Patterned solution to the model equations (\ref{eqn: the model}). Savanna trees formed regular spots within a grassy matrix, with slightly higher grass biomass under the tree canopies due to shading. In line with the multistability exhibited by ecosystem models \citep{Rietkerk2021}, variations in initial conditions may drive the system to various distinct patterns (typically with different spatial periods and configuration).}
\label{standard_grass_tree_patterns}
\end{figure}

In this final state, interacting localized structures form a large-amplitude, spatially periodic pattern characterized by steep gradients, typically induced by differences in the diffusion rates $d_g$ and $d_s$. These vegetation patterns, an example of which is illustrated in Figure \ref{standard_grass_tree_patterns} (for the numerical method and the specific parameter values see Appendix \ref{numerical method}), feature patches of savanna trees embedded within a grassy landscape. Grass biomass is slightly higher under tree canopies, benefiting from the shading effect that reduces moisture stress. The distribution, size and shape of these patches vary sensitively with initial conditions and parameters. These findings resemble observations of spatially periodic patterning originating from subcritical Turing bifurcations in earlier vegetation models \citep{Rietkerk2002,Tzuk2020}. 

\subsection*{Turing-evades-tipping}

\begin{figure}[h!]
    \centering
    \begin{adjustbox}{center, margin=0cm}
  \includegraphics[width=\textwidth]{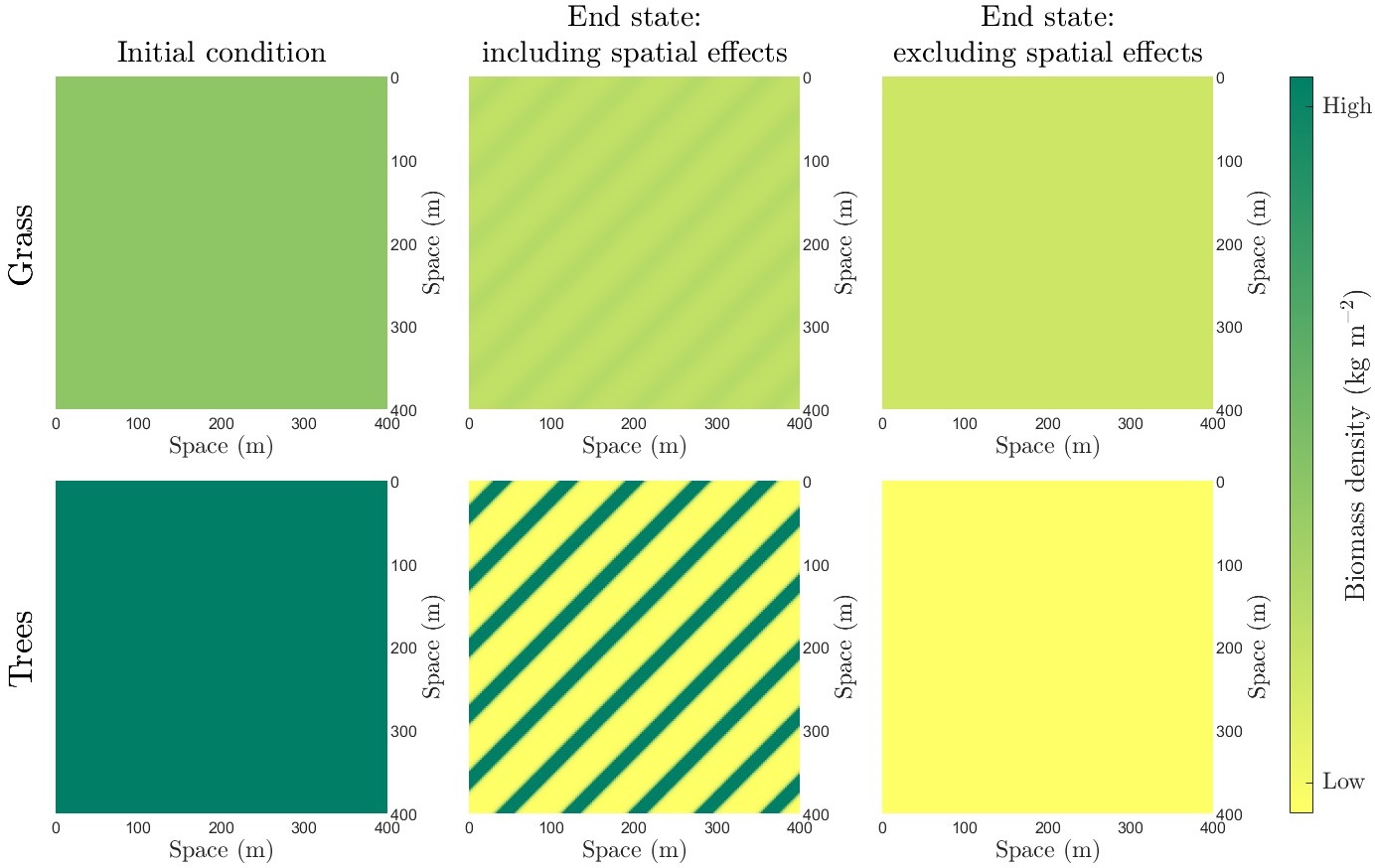}
\end{adjustbox}
\caption{Simulations of the model equations (\ref{eqn: the model}) in both spatial ($d_g,d_s>0$) and non-spatial settings ($d_g=d_s=0$) illustrating the Turing-evades-tipping mechanism. Starting from a uniform mixed tree-grass state, the spatial model exhibits persistent vegetation stripes, while the non-spatial model tips into a grassland state with lower biomass density as $a_s$ is slowly decreased beyond the tipping point $a_{s,SN}$.}
\label{Turing_evades_tipping}
\end{figure}

Having established the existence of patterned solutions, we now examine how these vegetation patterns influence ecosystem resilience. Specifically, we assess whether the Turing-evades-tipping mechanism, where spatial pattern formation prevents ecosystem collapse, manifests in our model. Our findings confirm the presence of this mechanism (Figure \ref{Turing_evades_tipping}). Through simulations, we compare the spatial and non-spatial versions of the model, both initialized from a mixed state of trees and grasses. The uniformly mixed state is stable in the non-spatial model, but in the spatial model, vegetation stripes emerge. As the bifurcation parameter $a_s$ is slowly decreased beyond the saddle-node bifurcation at $a_{s,SN}$, the non-spatial model tips and undergoes a critical transition, leading to the collapse of tree cover and a reduction in grass biomass, resulting in a treeless grassland state. In contrast, the patterned state in the spatial model persists, demonstrating resilience despite the same parameter shift.

\subsection*{Turing-triggers-tipping}

\begin{figure}[t!]
    \centering
    \begin{adjustbox}{center, margin=0cm}
  \includegraphics[width=\textwidth]{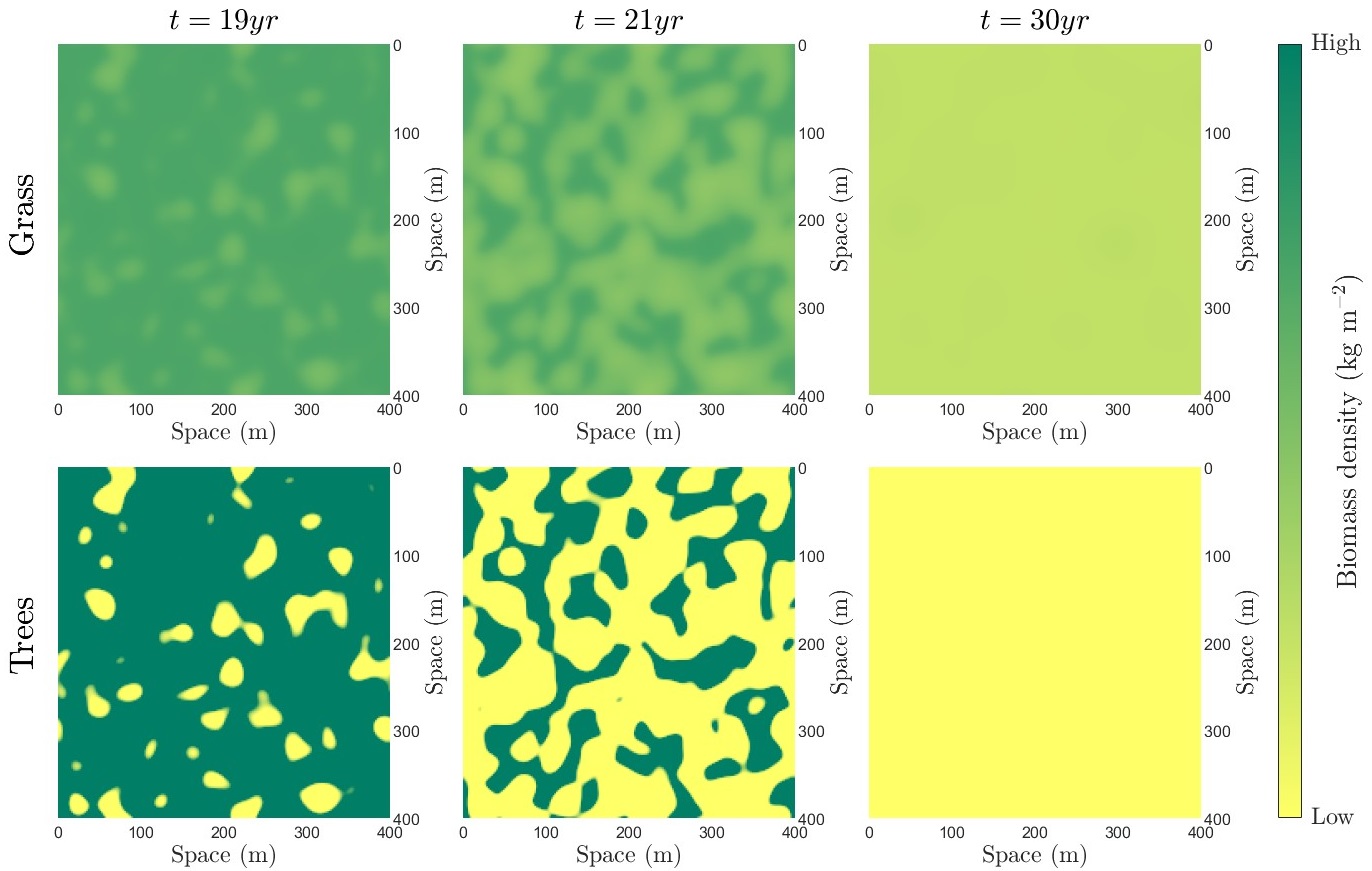}
\end{adjustbox}
\caption{Time-lapse of the Turing-triggers-tipping mechanism for $a_{s,SN}<a_s<a_{s,T}$. The system transitions from a savanna state to a lower-biomass grassland state via a patterned transient. In the non-spatial model, tipping occurs only after the saddle node bifurcation, allowing the savanna state to remain stable for $a_s>a_{s,SN}$.}
\label{Transient_Turing}
\end{figure}

In the previous sections, we showed that a subcritical Turing bifurcation can drive the system towards large amplitude spatially periodic patterns. However, we can observe other model outcomes for different parameter settings. Instead of stabilizing into a patterned solution, the system may transition to a homogeneous equilibrium, in particular the grassland state, with spatial patterns acting as a transient (Figure \ref{Transient_Turing}). Initially, patches of treeless grassland emerge across the savanna landscape. Over time, the tree-dominated patches are gradually overtaken by grasses, leading the system to converge to a homogeneous, lower-biomass grassland state in a process that resembles gradual tipping or regime shift proceeding by front propagation \citep{Bel2012,Zelnik2018}. This patterned transient highlights the instability of the savanna state, which fails to persist and undergoes a critical transition instead.

We define this phenomenon as ``Turing-triggers-tipping". In the non-spatial model, tipping occurs only after crossing the saddle-node bifurcation at $a_{s,SN}$, resulting in a rapid transition. In contrast, in the spatial model, tipping can be triggered earlier, within the range $a_{s,SN}<a_s<a_{s,T}$, following the Turing bifurcation. Here, the emergence of patterns signals the onset of collapse rather than ecosystem stabilization. This suggests that, contrary to recent literature associating pattern formation with ecosystem resilience \citep{Rietkerk2021}, spatial patterns may actually also drive ecosystem collapse under environmental stress.

Our simulations also reveal differences in the timescales of the tipping events. Turing-triggers-tipping unfolds gradually, typically over several decades. In contrast, classical tipping via the saddle-node bifurcation, a non-spatial mechanism, is much faster, often occurring within a few years. The extended timescale of tipping through transient patterning offers a crucial window for early detection of ecosystem collapse \citep{Zelnik2018}. Notably, Turing-triggers-tipping is characterized by irregular transient patterns and the gradual expansion of tree-free patches, which may help distinguish it from Turing-evades-tipping in empirical observations.

Systematic variation of individual parameters suggests that Turing-evades-tipping and Turing-triggers-tipping occur in distinct, non-overlapping regions of the parameter space, with no straightforward parametric relationships between them. Given the high dimensionality of the parameter space, determining a clear division of the parameter space into regions where each mechanism occurs is a complex task beyond the scope of the present paper.

In summary, while vegetation patterning under environmental stress may indicate ecosystem resilience following the Turing-evades-tipping framework, the Turing-triggers-tipping mechanism demonstrates that patterns can also serve as indicators of impending savanna collapse. The final outcome, whether the system converges to a patterned state or a homogeneous state, depends on the specific parameter setting and initial condition. The extended transient phase associated with Turing-triggers-tipping offers a key opportunity to detect early warning signs of ecosystem instability, providing valuable insights for assessing ecosystem resilience under changing environmental conditions.

\section{Discussion}

In this paper, we introduced a minimalistic spatially extended model of tree-grass dynamics in dry savannas, incorporating key recommendations from the comprehensive review by \citep{Holdo2023}. The model captures the asymmetry between competition and facilitation across tree life stages: grasses inhibit tree establishment by monopolizing surface water, but mature trees, with deeper and flexible root systems can overcome this limitation and additionally alleviate moisture stress for grasses through shading. To simplify the model, soil moisture is treated implicitly through interaction terms and parameter dependence, rather than as an explicit variable.

Our non-spatial analysis shows that these mechanisms induce stable tree-grass coexistence, supporting the hypotheses of \citep{Holdo2023}. When spatial effects are included, facilitation drives vegetation pattern formation under environmental stress. This patterning may enhance resilience through the Turing-evades-tipping framework or, conversely, trigger collapse via the Turing-triggers-tipping mechanism. The specific outcome, whether stabilization or transition, depends on parameters and initial conditions, though the underlying determinants remain uncertain. Future studies should quantify the likelihood of each process and identify parametric relationships governing shifts between these mechanisms through further mathematical analysis.

\subsection*{A novel perspective on ecosystem transitions}

A key contribution of this work is identifying transient Turing patterns as drivers of ecosystem transitions, a phenomenon we term ``Turing-triggers-tipping". In this framework, spatial instabilities lead to pattern formation, but instead of stabilizing, the pattern amplitude continues to grow, ultimately driving a shift to another system state. In our model, this shift can push the system from a savanna state to a treeless grassland, resulting in the loss of critical ecosystem functions like habitat provision. This dynamic highlights that subcritical Turing bifurcations can trigger tipping events even before a critical threshold is reached in the non-spatial model. 

This finding suggests that the idea that spatial patterning inherently enhances savanna ecosystem resilience may be less common than previously thought \citep{Rietkerk2021}. While vegetation patterns have been considered signs of self-organizing, healthy systems \citep{Banerjee2024}, our results show that spatial effects can induce tipping earlier than predicted by non-spatial analyses, suggesting that spatial interactions may increase savanna ecosystem vulnerability in some cases. This insight emphasizes the need for a deeper understanding of the role that spatial dynamics play in ecological stability.

Although transient Turing patterns have been explored in models of neural networks \citep{Elvin2009}, predator-prey systems \citep{Xiao2023}, and morphogenesis \citep{Guisoni2022}, their relevance to vegetation dynamics is novel. The possibility of transient patterning induced by a subcritical Turing bifurcation is well-documented in mathematical studies \citep{Doelman2019}, and such phenomena appear more frequently in systems with multistability of homogeneous equilibria \citep{Krause2024}.

\subsection*{Towards a deeper understanding of ecosystem dynamics}

Understanding how stable and transient Turing patterns manifest in natural ecosystems is crucial for developing preemptive conservation strategies. The Turing-triggers-tipping mechanism observed in our dry savanna model indicates that spatial biomass heterogeneity could play a pivotal role in detecting ecosystem tipping points. The gradual dynamics of transient patterns provide a window for early intervention, contrasting with the sudden transitions typically associated with non-spatial tipping points. However, the slower pace of these dynamics makes it difficult to distinguish between transient patterns that drive tipping and stable patterns that prevent it. Thus, future studies should focus on differentiating between resilience-enhancing patterning (Turing-evades-tipping) and early warnings of collapse (Turing-triggers-tipping). Notably, patterns resulting from Turing-triggers-tipping are expected to be more irregular, with a gradual expansion of tree-free patches, which could aid in distinguishing them from Turing-evades-tipping. Given uncertainties around the observability of these patterns in natural settings, we emphasize the need for further empirical research.

The nature of the Turing bifurcation, whether supercritical or subcritical, significantly influences system behaviour, as only subcritical bifurcations produce transient Turing patterns. Distinguishing between these forms by deriving the Landau coefficient provides critical insights into the range of possible dynamics \citep{Doelman2019}. Additionally, further investigation of the patterned solutions in our model offers a promising direction for future work. The low survival rates of early-life stage trees under harsh conditions \citep{Wakeling2015}, represented in our model by the Allee effect, suggest that spatial patterns may commonly form in dry savannas. As for the analysis of dryland models \citep{Bastiaansen2019,Bastiaansen2020}, advanced mathematical tools such as Geometric Singular Perturbation Theory \citep{Hek2010}, could aid in analyzing these complex coexistence states. 

Furthermore, even without a Turing bifurcation, the spatial model could introduce significant differences with the non-spatial model. Whereas the non-spatial model ultimately leads to dominance by a single state (savanna or grassland, depending on initial conditions), the spatial model supports diverse configurations. These potentially include stable coexistence fronts separating grassland and savanna as well as more intricate spatial structures, such as front instabilities \citep{Fernandez2019, Carter2023, Banerjee2024}.

Our model captures the key interactions driving tree-grass dynamics in dry savannas, as outlined by \citep{Holdo2023}. Most of these mechanisms are water-mediated, and future work could include water as an explicit variable for a more realistic and mechanistic modelling effort, which could possibly be better parameterized from observations. This would offer a clearer understanding of how water availability influences ecosystem stability and collapse, addressing potential limitations of our simplified model. Moreover, incorporating other ecological factors, such as fire regimes and herbivory (more critical in mesic savannas \citep{Sankaran2005,DOnofrio2018}), adding complex non-local seed dispersal mechanisms \citep{Patterson2024} or considering phenotypic transitions \citep{Bennett2023}, would further enhance our understanding of how spatial patterning impacts long-term savanna dynamics.

In summary, our model successfully captures tree-grass coexistence in dry savannas, already in the non-spatial case, while the spatial model highlights the critical role of spatial interactions and facilitation in driving ecosystem resilience and transitions. We demonstrate how both the Turing-evades-tipping framework and the Turing-triggers-tipping mechanism influence resilience and collapse, offering new insights into the complex dynamics of savanna ecosystems under environmental stress.

\section{Acknowledgments}
This research was supported by the Dutch Research Council (NWO) through the project ``Resilience in Complex Systems through Adaptive Spatial Pattern Formation'' (project number OCENW.M20.169), with contributions from J.V., M.R., and A.D. Additional funding for E.M., M.R., and A.D. was provided by the European Research Council through the ERC-Synergy project RESILIENCE (proposal number 101071417). M.B. acknowledges support from the Italian National Biodiversity Future Center (NBFC): National Recovery and Resilience Plan (NRRP), Mission 4 Component 2 Investment 1.4 of the Italian Ministry of University and Research; funded by the European Union – NextGenerationEU (project code CN00000033). We extend our gratitude to Paul Zegeling and Paul Carter for their advice in developing the numerical code used in this research.

\newpage

\newpage

\begin{appendices}

\section{Parametrization and calibration}
\label{parametrization and calibration}

This appendix outlines the parameterization and calibration process used to derive ecologically plausible ranges for the model parameters. Our model simplifies tree-grass interactions not explicitly modelling water redistribution, instead focusing on facilitation and competition between grasses and trees across tree life stages, and the role of spatial interactions. This simplification makes it challenging to directly obtain parameter values from empirical data, which is a common limitation of theoretical models. Therefore, we follow standard practice of deriving realistic parameter ranges from literature rather than pinpointing exact values, capturing a wide spectrum of qualitative mechanisms driving ecosystem dynamics across the parameter space. This approach allows us to explore whether the mechanisms we are studying lead to stable and resilient coexistence in savannas, and examine how spatial effects influence model outcomes compared to non-spatial models.

Below, we outline the parameterization process and the ranges identified. For the parameters $a_s$, $p_f$ and $d_s$, we used calibration (see descriptions at the end).

\subsubsection*{Parametrization}

\begin{itemize}
    \item $m_g$ (Mortality rate of grasses): The lifespan of grasses typically range from 1 to 3 years \citep{Francesco2010}. Disregarding both growth and facilitation, the grass biomass decreases exponentially with a factor of $\left(e^{-m_g}\right)^x$ after $x$ years. Assuming one percent of the grasses remain after 1 and 3 years, we calculate $m_g = \ln{100}$ (for 1 year) and $m_g = \ln{100}/3$ (for 3 years). Therefore, we define the range as $\ln{100}/3 < m_g < \ln{100}$.
    \item $a_g$ (Growth rate of grass): This parameter governs the exponential growth rate of grasses in absence of trees and far away from the maximum standing biomass. Data suggests that the relative growth rate of grasses can reach values up to $0.3$ yr\textsuperscript{-1} \citep{simpson2023}. To account for variability, we set an upper bound of $0.5$, resulting in $1 < e^{a_g-m_g} < 1.5$. Thus, we find $m_g < a_g < m_g + \ln{1.5}$ yr\textsuperscript{-1}.
    \item $K_g$ (Theoretical grass biomass cap): According to \citep{Moustakas2013}, grass biomass densities further away from the tree canopy in drier sites typically range between $0.2$ and $0.5$ kg m\textsuperscript{-2}. Based on this observation, we impose the condition $0.2 < g_{cc} < 0.5$ kg m\textsuperscript{-2}, where $g_{cc} = K_g\left(1-\frac{m_g}{a_g}\right)$ represents the grass carrying capacity in absence of trees. Considering the parameter ranges for $a_g$ and $m_g$, we found the range $0.2 < K_g < 2$ kg m\textsuperscript{-2} to adequately capture this constraint.
    \item $K_s$ (Theoretical tree biomass cap): Based on biomass estimates of savanna trees across precipitation gradients in dry savannas, tree biomass densities have been estimated as $2.8$ kg m\textsuperscript{-2} in a tree savanna and as $5$ kg m\textsuperscript{-2} in an open woodland \citep{fao1997}. Therefore, we impose the condition $2.8 < \tilde{s} < 5$ kg m\textsuperscript{-2}, where $\tilde{s}$ represents the carrying capacity for savanna trees in absence of grasses, as introduced in Appendix \ref{linear stability analysis}. Since $K_s$ represents an upper biomass limit, we extend the upper bound and define the range as $2.8 < K_s < 10$ kg m\textsuperscript{-2}.
    \item $C_s$ (Threshold tree density of Allee-effect): This parameter represents the critical threshold biomass density for early-life stage tree survival under dry conditions and should be significantly lower than $K_s$. We estimate $C_s < K_s/100$, with smaller values being more plausible in drier conditions.     
    \item $m_s$ (Mortality rate of trees): Savanna tree lifespans typically range from 10 to 100 years \citep{Francesco2010}. Following a similar approach to the parametrization of $m_g$, this results in the range $\ln(100)/100 < m_s < \ln(100)/10$.
    \item $p_c$ (Rate of competition): When no mature trees are present, $s$ is small, so $\beta(s) \approx 1$. At the grass carrying capacity $g_{cc}$, competition reduces to $-p_c g_{cc} s$. Early-life stage trees have very low survival rates due to intense competition with grasses \citep{Sankaran2004,Riginos2009}, so we assume $95$ to $99.9$ perish within a year. This gives the bounds $\ln(20)/g_{cc}<p_c<\ln(1000)/g_{cc}$ which simplifies to $2\ln(20)<p_c<5\ln(1000)$ based on the extremes of $g_{cc}$.
    \item $d_s$ (Diffusion rates of trees):     
    In our model, diffusion captures different ecological mechanisms, such as seed dispersal, germination and seedling establishment. We assume that tree diffusion rates are an order of magnitude smaller than those of grasses, reflecting the harsh conditions trees encounter during germination and establishment in dry savannas \citep{Wakeling2015}. Therefore, we set $d_s = d_g/10$.
    \item $s_f$ (Characteristic parameter of $\alpha(s)$): At the tree carrying capacity ($s = \tilde{s}$), $\alpha(s)$ should be relatively close to its maximum. Quantifying ``close'', we set this to at least $80$ percent of $1$, yielding the range given by $s_f = \tilde{s} \cdot \left(\frac{1}{F}-1\right)^{\frac{1}{n}}$ for $F \in [0.8,1)$.
    \item $s_c$ (Characteristic parameter of $\beta(s)$): At the tree carrying capacity ($s = \tilde{s}$), $\beta(s)$ should be relatively close to $0$. Quantifying ``close'', we set this to at most $20$ percent of $1$, yielding the range given by $s_c = \tilde{s} \cdot \left(\frac{C}{1-C}\right)^{\frac{1}{n}}$ for $C \in (0,0.2]$.
\end{itemize}

\subsubsection*{Calibration}
\begin{itemize}
    \item $a_s$ (Growth rate of trees): This parameter controls savanna tree growth. Over the gradient of this parameter we expect to see the transition between savanna and grassland. Calibration yields the range $0.001 < a_s < 0.1$ yr\textsuperscript{-1}.
    \item $p_f$ (Rate of facilitation): This parameter strongly influences the increase in grass biomass under the tree canopy due to shading effects. Based on \citep{Moustakas2013}, this facilitation should not exceed $100$ percent. Model calibration suggest the range $0 < p_f < 2$ yr\textsuperscript{-1}.
    \item $d_g$ (Diffusion rates of grasses): Based on model calibration, we set $10 < d_g < 100$ m\textsuperscript{2} yr\textsuperscript{-1}, ensuring realistic pattern wavelengths for a $400$ m x $400$ m spatial grid, in line with observed scales of patterning \citep{Groen2007}. 
\end{itemize}

\section{Linear stability analysis}
\label{linear stability analysis}

In this appendix, we outline the linear stability analysis used to generate the bifurcation diagrams in Figures \ref{Bifurcation diagram nonspatial} and \ref{Bifurcation diagram spatial}. This analysis allows us to assess the stability of the homogeneous equilibria, examining how these steady states respond to uniform and spatially varying perturbations.

\subsection*{Homogeneous steady states}

The homogeneous equilibria of the model equations (\ref{eqn: the model}) are determined by setting $\frac{\partial g}{\partial t} = \frac{\partial s}{\partial t} = 0$, resulting in multiple uniform steady states $(g_{ss}, s_{ss})$ for grass and tree biomass:
\begin{itemize}
    \item Bare soil state: $(0,0)$, representing the absence of vegetation.
    \item Grass-only state: $(K_g(1-\frac{m_g}{a_g}),0)$, which is ecologically relevant when $a_g>m_g$, ensuring sustainable grass growth.
    \item Tree-only states: $(0,\tilde{s})$, where $\tilde{s}$ is a solution to the quadratic equation:
    \begin{align}
        0=a_s\left(\frac{\tilde{s}}{C_s}-1\right)\left(1-\frac{\tilde{s}}{K_s}\right)-m_s, 
    \end{align}
    providing up to two possible tree-only equilibria.
    \item Savanna states: $(\bar{g},\bar{s})$, where both trees and grasses coexist. The tree biomass $\bar{s}$ is found by solving:
    \begin{align}
        0 = a_s\left(\frac{\bar{s}}{C_s}-1\right)\left(1-\frac{\bar{s}}{K_s}\right)-m_s - p_c K_g\left(1+\frac{p_f \alpha(\bar{s}) - m_g}{a_g}\right) \bar{s} \beta(\bar{s}),
    \label{s equation}
    \end{align}
    while the grass biomass $\bar{g}$ is given by:
    \begin{align}
    \bar{g} = K_g\left(1+\frac{p_f \alpha(\bar{s}) - m_g}{a_g}\right),
    \end{align}
    In our numerical simulations, $\alpha(s)$ and $\beta(s)$ are rational functions of $s$, and we solve equation (\ref{s equation}) using the Matlab function roots. For the parameter range of interest (Table \ref{tab: GT model}), we find up to two ecologically relevant savanna states, where both tree ($\bar{s} > 0$) and grass ($\bar{g} > 0$) biomasses coexist.
\end{itemize}

\subsection*{Stability against homogeneous perturbations}
To assess the stability of the steady states against homogeneous perturbations (i.e., perturbations that do not vary in space, identical to the stability analysis of the non-spatial model), we substitute $(g,s) = (g_{ss},s_{ss})+\varepsilon (\mu,\nu) e^{\lambda t}$ into the model and expand to leading order in $\varepsilon$. This results in the eigenvalue problem for the Jacobian matrix, evaluated at $(g_{ss}, s_{ss})$, given by
\begin{align}
    \mathcal{J}(g_{ss},s_{ss}) = \begin{pmatrix}
        a_g - 2\frac{a_g}{K_g}g_{ss}-m_g+p_f \alpha(s_{ss}) & p_fg_{ss}\alpha'(s_{ss})\\
        -p_cs_{ss}\beta(s_{ss}) & X(s_{ss})-p_cg_{ss}\left(\beta(s_{ss})+s_{ss}\beta'(s_{ss})\right)\end{pmatrix},
\end{align}
where $X(s) = -a_s+2a_s\left(\frac{1}{C_s}+\frac{1}{K_s}\right)s-3\frac{a_s}{C_sK_s}s^2-m_s$. The stability of a steady state $(g_{ss},s_{ss})$ is determined by the eigenvalues of this matrix. A steady state is stable if the real parts of all eigenvalues are negative, and unstable if any eigenvalue has a positive real part.

In our analysis, we find that the bare soil state is unstable, while the grass-only state is stable. Tree-only states are typically unstable in our numerical experiments. For the savanna states, stability varies across different parameter settings, with both stable and unstable states often observed simultaneously. This pairing frequently arises from a saddle-node bifurcation, as shown in the bifurcation diagrams of Figures \ref{Bifurcation diagram nonspatial} and \ref{Bifurcation diagram spatial}.

\subsection*{Stability against heterogeneous perturbations}
To analyze the stability of the uniform equilibria against spatially heterogeneous perturbations, we substitute $(g,s) = (g_{ss},s_{ss})+\varepsilon (\mu,\nu) e^{\lambda t+ikx} +\rm{c.c.}$, where $k$ is the wave number of the perturbation. Expanding this to leading order in $\varepsilon$, we derive the eigenvalue problem for the following matrix:
\begin{align}
    \begin{pmatrix}
         A - d_g k^2 & B\\
        C & D - d_s k^2
    \end{pmatrix},
\end{align}
where
\begin{align}
\begin{pmatrix}
        A & B\\
        C & D
    \end{pmatrix} \coloneqq \mathcal{J}(g_{ss},s_{ss}).
\end{align}
The corresponding dispersion relation is:
\begin{align}
\lambda^2 + \left(\left(d_g + d_s\right)k^2 - A - D\right) \lambda + d_g d_s k ^4 - (d_s A + d_g D)k^2 + AD - BC = 0
\end{align}
The solutions $\lambda_+(k^2)$ and $\lambda_-(k^2)$ to this quadratic equation determine the stability of the steady state under spatial perturbations. A Turing bifurcation occurs when one of these eigenvalues crosses zero (i.e., $\text{Re}(\lambda) = 0$), marking the onset of a spatial instability. If there is a range $(k_-^2, k_+^2)$ where one of the eigenvalues has a positive real part, the uniform steady state is spatially unstable, leading to the emergence of Turing patterns.

\begin{figure}[h!]
    \centering
    \begin{adjustbox}{center, margin=0cm}
  \includegraphics[width=0.6\textwidth]{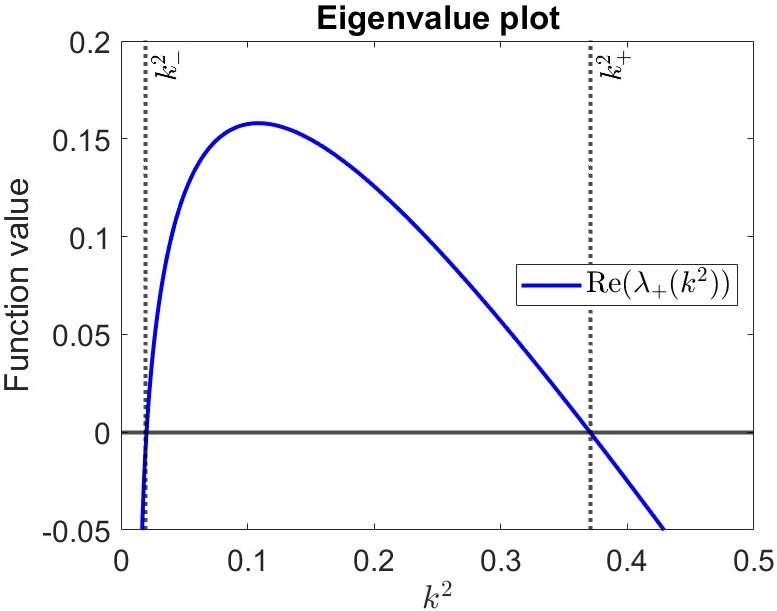}
\end{adjustbox}
\caption{Plot of the real part of the eigenvalue $\lambda_+(k^2)$ as a function of $k^2$, showing the range $(k_-^2, k_+^2)$ where $\text{Re}(\lambda_+(k^2)) > 0$. This corresponds to the spatial instability of the savanna state, leading to the development of transient Turing patterns as illustrated in Figure \ref{Transient_Turing}.}
\label{dispersion_relation}
\end{figure}

Our analysis indicates that the savanna state is the only steady state capable of undergoing a Turing bifurcation. In Figure \ref{dispersion_relation}, we plot the real part of $\lambda_+(k^2)$ for the savanna state, corresponding to a parameter combination where transient Turing patterns emerge (as shown in Figure \ref{Transient_Turing}). The plot shows that $\text{Re}(\lambda_+(k^2)) > 0$ for $k^2 \in (k_-^2, k_+^2)$, confirming the spatial instability of the uniform savanna state.

\section{Numerical setting}
\label{numerical method}

This appendix details the numerical methods used to produce the model simulations presented in Figures \ref{standard_grass_tree_patterns}, \ref{Turing_evades_tipping}, and \ref{Transient_Turing}. Additionally, we specify the functional forms of $\alpha(s)$ and $\beta(s)$ and list the parameter choices that generated these figures, as well as the bifurcation diagrams in Figures \ref{Bifurcation diagram nonspatial} and \ref{Bifurcation diagram spatial}.

All simulations were conducted using Matlab. We discretized the two-dimensional spatial domain using the Method of Lines, which converts partial differential equations into a system of ordinary differential equations \citep{Schiesser2009}. Temporal integration was carried out using Matlab’s ode15s solver, which is optimized for stiff systems, ensuring stability and efficiency in simulations that involve both fast and slow dynamics \citep{MATLABode15s}. The initial conditions consisted of a homogeneous savanna state, perturbed by small fluctuations to introduce spatial heterogeneity. Periodic boundary conditions were applied to simulate an effectively infinite domain, minimizing boundary effects and accurately reflecting the interior dynamics of a large savanna ecosystem.

For the interactions between trees and grasses, we employed the following functional forms for $\alpha(s)$ (facilitation by trees) and $\beta(s)$ (competition from grasses): 
\begin{align} 
\alpha(s) &= \frac{\left(s/s_f\right)^n}{1 + \left(s/s_f\right)^n}, \\ 
\beta(s) &= \frac{1}{1 + \left(s/s_c\right)^n}. 
\end{align} 
These forms satisfy the necessary conditions that $\alpha(0) = \lim_{s \to \infty} \beta(s) = 0$ and $\beta(0) = \lim_{s \to \infty} \alpha(s) = 1$, ensuring proper behaviour at extreme values of biomass density. The chosen functional forms also simplify the numerical computation of the mixed equilibrium states and generate the desired sigmoid shape, consistent with the dynamics shown in Figure \ref{fig: Tree-Grass Interaction Dynamics}. For all simulations, we set $n = 2$. The parameters $s_f$ and $s_c$ were selected so that at $s = \tilde{s}$, the functions $\alpha(s)$ and $\beta(s)$ are relatively close to their maximum and minimum, respectively.

The parameter values used in the simulations and bifurcation analyses are presented in Table \ref{tab: parameter simulations}, which are consistent with the ecologically relevant ranges (see Table \ref{tab: GT model}).

    \begin{table}[t!]
\centering
\arrayrulecolor{black}
\renewcommand{\arraystretch}{1}
\footnotesize
\begin{tabular}{|>{\centering\arraybackslash}m{1.42cm}|>{\centering\arraybackslash}m{1.42cm}|>{\centering\arraybackslash}m{1.42cm}|>{\centering\arraybackslash}m{1.42cm}|>{\centering\arraybackslash}m{1.42cm}|>{\centering\arraybackslash}m{1.42cm}|>{\centering\arraybackslash}m{1.42cm}|}
\hline
\rowcolor[HTML]{EFEFEF} 
\textbf{Param} & \textbf{Bif I} & \textbf{Bif II} & \textbf{Sim I} & \textbf{Sim II} & \textbf{Sim III} & \textbf{Units} \\ \hline
$a_g$ & $1.95$  &  $2.37$  &  $1.95$  &  $1.86$  &  $3.19$  & yr\textsuperscript{-1} \\ \hline
$K_g$ &  $1.99$  &  $1.59$  &  $1.43$  &  $1.90$  &  $1.93$  & kg m\textsuperscript{-2} \\ \hline
$m_g$ &  $1.56$  &  $2.01$  &  $1.64$  &  $1.65$  &  $2.85$  & yr\textsuperscript{-1} \\ \hline
$p_f$ &  $0.178$  &  $0.228$  &  $0.408$  &  $0.291$  &  $0.277$  & yr\textsuperscript{-1} \\ \hline
$d_g$ &  $100$  &  $100$  &  $100$  &  $100$  &  $50$  & m\textsuperscript{2} yr\textsuperscript{-1} \\ \hline
$a_s$ &  $0.0293$  &  $0.0621$  &  $0.0365$  &  $0.00962$  &  $0.0111$  & yr\textsuperscript{-1} \\ \hline
$C_s$ &  $0.0368$  &  $0.0229$  &  $0.00837$  &  $0.00442$  &  $0.00342$  & kg m\textsuperscript{-2}   \\ \hline
$K_s$ &  $4.05$  &  $5.02$  &  $3.87$  &  $4.85$  &  $3.98$  & kg m\textsuperscript{-2}  \\ \hline
$m_s$ &  $0.268$  &  $0.246$  &  $0.0917$  &  $0.135$  &  $0.221$  & yr\textsuperscript{-1} \\ \hline
$p_c$ &  $9.37$  &  $23.6$  &  $24.4$  &  $16.4$  &  $30.5$  & yr\textsuperscript{-1} \\ \hline
$d_s$ &  $10$  &  $10$  &  $10$  &  $10$  &  $5$  & m\textsuperscript{2} yr\textsuperscript{-1} \\ \hline
$n$ &  $2$  &  $2$  &  $2$  &  $2$  &  $2$  & - \\ \hline
$s_f$ &  $1.44$  &  $2.28$  &  $1.52$  &  $2.08$  &  $1.66$  & kg m\textsuperscript{-2} \\ \hline
$s_c$ &  $0.789$  &  $2.39$  &  $1.82$  &  $2.20$  &  $1.60$  & kg m\textsuperscript{-2} \\ \hline
\end{tabular}
\caption{Parameter settings used to generate the figures in this paper, based on the ecological ranges shown in Table \ref{tab: GT model}. The columns ``Bif I'' and ``Bif II'' correspond to the bifurcation diagrams in Figures \ref{Bifurcation diagram nonspatial} and \ref{Bifurcation diagram spatial}, respectively. The columns ``Sim I'', ``Sim II'' and ``Sim III'' correspond to model simulations illustrated in Figures \ref{standard_grass_tree_patterns}, \ref{Turing_evades_tipping} and \ref{Transient_Turing}, respectively.}
\label{tab: parameter simulations}
\end{table}

\end{appendices}

\newpage

\bibliographystyle{abbrvnat}
\bibliography{ref}

\end{document}